# Spin-Coupled Local Distortions in Multiferroic Hexagonal $HoMnO_3$

T. A. Tyson[1,2], T. Wu[1], K. H. Ahn[1,2], S. Kim[2] and S.-W. Cheong[2]
[1]Department of Physics, New Jersey Institute of Technology, Newark, NJ 07102
[2]Department of Physics and Astronomy and Rutgers Center for Emergent Materials, Rutgers University, Piscataway, NJ 08854

## Abstract

Local structural measurements have been performed on hexagonal $HoMnO_3$ in order to ascertain the specific changes in bond distances which accompany magnetic ordering transitions. The transition from paramagnetic to the antiferromagetic (noncollinear) phase near ~70 K is dominated by changes in the a-b plane Mn-Mn bond distances. The spin rotation transition near ~40 K involves both Mn-Mn and nearest neighbor Ho-Mn interactions while the low temperature transition below 10 K involves all interactions, Mn-Mn, Ho-Mn (nearest and next nearest) and Ho-Ho correlations. These changes in bond distances reveal strong spin-lattice coupling. The similarity in magnitude of the change in J(Mn-Mn) and J(Ho-Mn) enhances the system frustration. The structural changes are interpreted in terms of a model of competing spin order and local structural distortions. Density functional calculations are used to estimate the energies associated with ionic displacements. The calculations also reveal asymmetric polarization of the charge density of Ho, O3 and O4 sites along the z-axis in the ferroelectric phase. This polarization facilitates coupling between Ho atoms on neighboring planes normal to the z-axis.

## I. Introduction

The existence and coupling of distinctly different types of long range order is stimulating much interest from a fundamental science perspective. The rare earth hexagonal manganites corresponding to systems with the chemical formula $REMnO_3$ (RE ions with small radii such as Sc, Y, Ho, Er, Tm, Yb, and Lu) exhibit both ferroelectric and magnetic order. In this specific class of materials the transition to the ordered ferroelectric state ($T_{FE}$) occurs between ~800 and ~1200 K while the ordered magnetic states occur at significantly lower temperature ($T_N$~80 K). This hexagonal structure can also be stabilized in large radius cation systems by quenching from high temperature or by depositions on substrates which induce strain on the samples [1, 2, 3, 4, 5, 6, 7].

In the hexagonal structure, the $Mn^{3+}$ ions from a triangular planar lattice where the Mn spins are coupled indirectly through the planar oxygen atoms. Locally the Mn ions are incorporated into $MnO_5$ polyhedra (bipyramids) with three oxygen atoms surrounding Mn in a plane and two apical oxygen atoms above and below the plane completing the polyhedron. Excluding the oxygen atoms, the Mn ions form a triangular two-dimensional lattice, with each two dimensional layer separated by RE site ions.

The degree of frustration in these systems, indicated by the ratio of the Curie-Weise constant (derived from $1/\chi$) to the Neel temperature, approaches 10 in the hexagonal systems [4]. The frustration is relieved by transition into a unique in-plane magnetic ordering in which the Mn ion moment are pointed $120^{\circ}$ relative to each other. In the systems with open 4f shells ($REMnO_3$), a series of magnetically ordered states are found since, it is believed, that the spins on the RE ions can both couple with the Mn ions and each other and order magnetically and at low temperatures [6,7]. In $HoMnO_3$ for example, neutron diffraction measurements reveal a $90^{\circ}$ in-plane rotation of the Mn spins near ~ 40 K ($T_{SR}$) followed by magnetic ordering of the Ho sites below ~ 5 K ($T_{Ho}$).

Anomalies in the dielectric constants, the linear expansion coefficients and phonon frequencies suggest a coupling between the magnetic and ferroelectric order at low temperature [8, 9]. In $HoMnO_3$, sharp features are observed at $T_{SR}$ and $T_{Ho}$. Heat capacity and magnetic susceptibility measurements in magnetic field reveal that while the position of $T_N$ is stable with respect to external magnetic fields the $T_{SR}$ transition temperate is strongly field dependent and moves to lower temperature with increasing magnetic fields [5]. Moreover, a complex multiphase structure is found at low temperature. In magnetization studies on $HoMnO_3$ no slope changes are found at $T_N$. However, changes in slope were found at $T_{SR}$ and $T_{HO}$. The abrupt changes in the c-axis magnetization at $T_{SR}$ were attributed to changes in the magnetic order of the $Ho^{3+}$ ions since the Mn spin is confined to the a-b plane. A lambda type anomaly occurs in the heat capacities at $T_N$ in zero magnetic field. The peaks at $T_{SR}$ and $T_{Ho}$ are found to exhibit sharp onsets suggesting first order transitions. A complex phase diagram is revealed in the H-T plane at low temperature.

Under pressure, the sharp peak in the dielectric constant seen at $T_{SR}$ moves to lower temperature and is reduced in amplitude as pressure is increased from 0.1 GPa to 1.7 GPa [8]. In addition, the a-b plane lattice parameter is more readily compressed with pressure than the c-axis lattice parameter [10]. Hence, the in plane Mn-Mn correlation may be preferentially modified by pressure. This suggests that strain may suppress the magnetic transitions in $HoMnO_3$ and other hexagonal $RMnO_3$ systems. Indeed, as a result of tensile strain, the magnetic structure of c-axis oriented epitaxial $HoMnO_3$ films differs from that of bulk system and has a significantly reduced $T_N$ (~20 K lower) [11, 12].

In the low temperature ferroelectric phase, it is argued that there is a tilting of the $MnO_5$ polyhedra towards O(3) at the threefold rotation center and a buckling of the RE plane [13]. RE is off the center of two in-plane oxygen atoms along the c-axis and this produces a dipole along the c-axis in one O-RE-O-RE chain along the c axis. The number of RE(1) chains and RE(2) chains are not equal and hence a net dipole along c is produced. The low temperature moment on the Mn site is found to vary between ~3 $\mu_B$ and ~3.5 $\mu_B$ (not 4 $\mu_B$ expected for a $Mn^{3+}$ ion) possibly due to the persistence of strong fluctuations even below the ordering temperature of the RE site. In $LuMnO_3$ strong magnetic diffuse scattering exist up to approximately $3T_N$ indicating that spin fluctuation exist up to this temperature. Doping with Zr stabilizes the diffuse scattering down to very low temperatures suggesting short range antiferromagnetic clusters.

Recent work suggest that there is no rehybridization or change in the chemical bonding between the high temperature paraelectric and low temperature ferroelectric phases in $YMnO_3$ [2]. The Born effective charges are found to be close to the values expected based on the formal valence values with respect to the high temperature unit cell. A large Op displacement was shown to be composed of two components by group theoretical analysis. $K_3$ corresponds to a tilting of the Mn-O polyhedra leading to a tripling of the unit cell volume and a Γ-point component related to Y-Op displacements and non-cancelling polarization due to the buckling of the $MnO_5$ pyramids. Further analysis of the combined group theoretical and density functional analysis of $YMnO_3$ revealed a single mode ($K_3$) at the boundary couples to the polarization [14]. The structural changes resulting from softening of the $K_3$ mode induces a net polarization. The $K_3$ mode involves motion of Y at the 2a and 4b positions position (Y1 and Y2) and oxygen at the 6c, 2a and 4b positions (O1, O2, O3 and O4). The dominant displacements are for the Y1 and O3 ions along the c axis.

Neutron diffraction measurements reveal the long range order of the Ho moments onsets at $T_{SR}$ and that the ordered moment on the Mn site is enhanced below the $T_{Ho}$. This shows that the Mn and Ho spin sublattices are coupled. Measurement of the spin wave dispersion for the spins on the $Mn^{3+}$ gives good agreement with a two dimensional nearest neighbor Heisenberg model [3].

In this work, detailed temperature dependent local structural measurements were conducted on hexagonal $HoMnO_3$ in order to determine which bond pairs were associated with the changes in the dielectric constants at the magnetic ordering temperatures. The local structural measurements enable a microscopic level assessment of the spin-lattice coupling on the level of specific pairs of atoms. The transition from paramagnetic to the antiferromagnetic (noncollinear) phase near ~70 K ($T_N$) is dominated by changes in the a-b plane Mn-Mn bond distances. The spin rotation transition ($T_{SR}$) near ~40 K involves both Mn-Mn and nearest neighbor Ho-Mn interactions while the low temperature transition ($T_{Ho}$) below 10 K involves all interactions, Mn-Mn, Ho-Mn (nearest and next nearest) and Ho-Ho correlations. Density functional calculations reveal asymmetric polarization of the charge density of Ho, O3 and O4 sites along the z-axis in the ferroelectric phase.

## II. Experimental and Computational Methods

Polycrystalline samples of hexagonal $HoMnO_3$ were prepared by solid state reaction. X-ray absorption samples were prepared by grinding and sieving the material (500 mesh) and brushing it onto Kapton tape. Layers of tape were stacked to produce a uniform sample for transmission measurements with jump μt ~ 1. Spectra were measured at the NSLS beamline X23B at Brookhaven National Laboratory. Measurements were made on warming from 6 K to 300 K in a sample attached to the cold finger of a cryostat. Two to six scans were taken at each temperature. The uncertainty in temperature is < 0.2 K. At the Mn K-Edge, a Mn foil reference was employed for energy calibration. The reduction of the x-ray absorption fine-structure (XAFS) data was performed using standard procedures [15].

To treat the atomic distribution functions on equal footing at all temperatures the spectra were modeled in R-space by optimizing the integral of the product of the radial distribution functions and theoretical spectra with respect to the measured spectra. Specifically the experimental spectrum is modeled by, $\chi(k) = \int \chi_{th}(k,r)\, 4\pi r^2 g(r) dr$ where $\chi_{th}$ is the theoretical spectrum and g(r) is the real space radial distribution function based on a sum of Gaussian functions ($\chi(k)$ is measured spectrum) [16] at each temperature as in Ref. (as in Ref. [17]). Theoretical spectra for atomic shells [18] were derived from the room temperature crystal structure [7] shown in Fig. 1. For the Mn K-Edge, the k-range $2.54 < k < 11.5$ Å$^{-1}$ and the R-range $0.67 < R < 4.08$ Å was used while for the Ho L3-Edge the k-range $2.96 < k < 14.15$ Å$^{-1}$ and r-range $1.27 < R < 4.08$ Å were utilized. Coordination numbers for the atomic shells were fixed to the crystallographic values. For the Mn K-Edge and the Ho L3 edges $S_0^2$ (accounting for electron loss to multiple excitation channels) values were fixed at 0.80 and 0.96, respectively. The Gaussian widths and positions were fit for each component. Representative XAFS data and fits at 170 K are shown in Fig.2 and Fig. 3 for the Mn K-edge and Ho L3-Edge, respectively. The limited energy range at the Mn K-edge constrained our modeling to the shells: <Mn-O1, Mn-O2, Mn-O3>, Mn-O4, Mn-Mn and Mn-Ho (long, short). For the Ho L3 edge, modeling was restricted to the <Ho-O>, Ho-Mn(short), Ho-Ho and Ho-Mn(long) distances (see Fig. 1 for structural model).

To estimate the force constants spin density functional calculations in the projector augment wave approach [19] were carried out. Full optimization was conducted for both lattice parameters and atomic positions and the LDA+U approximation was implemented to obtain the relaxed structure. The ground state structure was optimized so that forces on each atom were below $7 \times 10^{-5}$ eV/Å. Self forces along x, y and z direction were computed for all atoms based on 0.03 Å displacements. Force constants were derived from combined positive and negative displacements.

To understand the charge distribution and bonding between the atoms in unit cell, calculation of the charge density within the local spin density approximation (U/J=0) was carried-out using the all-electron full potential linear augmented plane-wave plus local orbital (FPLAPW+lo) method implemented in the WIEN2k code [20]. The derived charge density plots were qualitatively insensitive to the use of finite U/J parameters.

## III. Results and Discussion

In Fig. 1(a) we show the structure of $HoMnO_3$ in the room temperature ferroelectric phase. Note that the Mn ions lie in a plane (Fig. 1(b)) forming a triangular lattice and are linked through O3 and O4 oxygen sites. Normal to this plane, the Mn ions are bonded to O1 and O2 apical oxygen atoms. We refer to the interactions between in-plane Mn ions (via the oxygen ions) as J1 exchange with an average Mn-Mn distance of ~3.5 Å and average Mn-O-Mn bond angle of ~ 119°. (Strictly speaking since there is not perfect triangular symmetry and hence there are multiple J1 interactions. However, here were are just attempting to rank the sizes of the various types of interactions). The Ho1 and Ho2 ions form a triangular plane which is buckled. The Ho sites interact with each other through the O1 and O2 apical oxygen sites below or above the buckled Ho plane. The average nearest neighbor Ho-Ho distance is ~3.6 Å with average Ho-O-Ho angles of ~ 102°. In addition to the Ho-Ho interactions and the Mn-Mn interactions there are also interactions between the Mn and Ho ions through the O1 and O2 apical oxygen ions. As a result of the bucking of the Ho planes the Ho-Mn bonds occur at two distinct distances which we call Ho-$Mn_{short}$ and Ho-$Mn_{long}$ with bond distances of ~3.3 and ~3.7 Å, respectively. The corresponding Mn-O-Ho bond angles are 106° and 125°, respectively. We label the long and short interactions as J3L and J3S. Even though the magnetic structure obtains a unique spin configurations at low temperature for the two-dimensional Mn triangular lattice, the presence of the Ho ions with localized 4f orbital ions will impact the magnetic structure at low temperature.

The J1 and J3S interactions should dominate at higher temperature due to the larger angles and/or shorter distances between the metal sites. This should be followed by the J3L and the and J2 interactions (See Fig. 7 for all interactions). We now look at the temperature dependence of the local structure to identify the specific bond changes associated with magnetic ordering temperatures.

The x-ray absorption measurements at the Mn K-Edge and Ho L3 edge enable us to examine the local structure about the Mn and Ho ions, respectively (Fig. 2 and Fig. 3). Temperature dependent structural analysis yield evidence for structural changes at magnetic ordering temperatures. In Fig. 4(a) we show the temperature dependence of the average Mn-O1, Mn-O2 and Mn-O3 bond distance ($<R_{Mn-O1}, R_{Mn-O2}, R_{Mn-O3}>$) giving mainly the out-of plane bond distance. At the level of the measurements no changes in out-of-plane Mn-O bond distances are seen. However, we note that the changes in the Mn-O3 bond are expected to anticorrelate with those of the Mn-O4 bond (as indicated by diffraction experiments) which we now discuss [4]. The in-plane Mn-O4 bond distance, on the other hand displays anomalies at all magnetic transitions $T_N$, $T_{SR}$ and $T_{Ho}$. , the largest change occurring at $T_N$. For comparison we show in Fig 4(b) the changes in the heat capacity (from Ref. [5]) at these temperatures revealing that they correspond to magnetic ordering transitions. Moreover, the anomalies in the dielectric constants (from Ref. [8]) at the magnetic ordering temperatures support the argument for strong spin-lattice coupling. However, here we are able to identify the specific atomic correlations associated with the magnetic ordering.

In Fig. 5(a) we show the temperature dependence of the average in-plane Mn-Mn distance (<Mn-Mn>) and average Ho-Ho distance (<Ho-Ho>). The <Mn-Mn> distance tracks the changes at all three magnetic ordering temperatures and is consistent with what is observed in the in-plane Mn-O4 distance mentioned above. On the other hand, no change is seen in the <Ho-Ho> distance except at $T_{Ho}$. In Fig. 5(b), the short and long Ho-Mn bond distances display intriguing behavior. While the long Ho-Mn bond exhibits (at most) a weak dip at $T_{SR}$, the short Ho-Mn bond exhibits a strong drop. At $T_{Ho}$ both bonds show sharp changes. For completeness we show the average Ho-O distance in Fig. 6. There is a suggestion of a small dip near $T_{SR}$ near the limit of detectability.

The measurements enable us to assess atomic origin of the magnetic ordering (Fig. 4(b)). The transition at $T_N$ is driven by strong in-plane Mn-Mn interactions (J1) supported by a reduction in the Mn-O4 and Mn-Mn distances. The transition at $T_{SR}$ is related both with the Mn-Mn in-plane correlations (J1) and the short Ho-Mn interactions (J3S). The transition at $T_{Ho}$, involves all interactions Mn-Mn (J1), Ho-Mn short (J3S), Ho-Mn long (J3L) and the Ho-Ho interaction (J2). We can thus estimate the relative sizes of the exchange interactions $|J1| > |J3S| > |J3L| > |J2|$. Fig. 7 shows all interactions discussed.

Recent measurements of the Ho magnetic ordering at low temperature on $HoMnO_3$ single crystals [21] reveal that both the Ho1 and Ho2 site order with spins along the c-axis and having antiferromagnetic coupling along c (J5). With this in mind, the structural changes involving the Ho ions can be understood in terms of a simple model of single-ion magnetic anisotropy [22]. Above the $Ho^{3+}$ ordering temperatures ($T_{Ho}$ and $T_{SR}$), the spins on the Ho sites are randomly arranged. Because of the high localization of the 4f electrons, the crystal field imposed by the oxygen neighbors does not modify the charge distribution of Ho atoms. An atomic picture can thus be used. The $Ho^{3+}$ ion has a highly anisotropic charge distribution with a prolate or pancake like squashed sphere. The magnetic moment on the $Ho^{3+}$ ions is rigidly coupled to the 4f charge density and is parallel to the symmetry axis. Because of the asymmetric shape of these ions rotation of the $Ho^{3+}$ moments must be accommodated by a change in the lattice. This explains the anomalies seen at $T_{SR}$ and $T_{Ho}$ in the local bond distances and in the dielectric constant (first order transitions). Moreover, the phonon modes will be altered as has been observed.

Returning to the origin of the spin rotation at $T_{SR}$, we note that the directions of spins in magnetically ordered states of transition metal oxides are usually determined by the spin-orbit interaction, or the magnetic anisotropy. The Mn and surrounding three O ions in the plane has the symmetry of isosceles triangles. The two apical Mn-O bond distances are not equal. Such low local symmetry splits all the Mn 3*d* orbitals, xy, yz, zx, $x^2$-$y^2$, and $3z^2$-$r^2$. With a strong Hunds coupling, Mn is expected to be in a high spin state, and the four 3d electrons occupy xy, yz, zx, and $x^2$-$y^2$ orbitals, since the average apical Mn-O bond distance is shorter than the average in-plane Mn-O bond distance. Therefore, the empty orbital state above and below $T_{SR}$ should remain identical, $3z^2$-$r^2$, and the spin-orbit interaction is not likely to change the spin orientation at $T_{SR}$. It is known that anisotropic exchange interaction, such as $H_{aniso,ex} = \sum \vec{S}_{\vec{i}}\ \hat{A}_{\vec{i},\vec{j}}\ \vec{S}_{\vec{j}}$, where $\hat{A}_{\vec{i},\vec{j}}$ is a 3 x 3 interaction matrix, plays an important role in multiferroic materials. For example, it is suggested that symmetric superexchange interaction and asymmetric Dzyaloshinskii-Moriya interaction contribute anisotropic exchange interaction between Ho and Mn ions in $HoMnO_3$ [1]. Unlike usual isotropic Heisenberg exchange interaction, anisotropic exchange interaction depends on the absolute spin orientations in addition to the relative spin orientation. The changes in local structure we have observed near $T_{SR}$ can influence the matrix elements in $\hat{A}_{\vec{i},\vec{j}}$ through magnetostriction effect, which can change the ground state spin orientation, giving rise to a spin rotation transition at $T_{SR}$. We also note that the changes in local structures are much more complex than the changes in average structures. It underscores the importance of the local probes for the study of multiferroic materials.

Using a simplified model, the change in bond distances at the magnetic ordering temperatures can be understood in terms of competing spin and lattice energies (see for example Ref [23]). The exchange energy J(x,y,z,θ) $\mathbf{S_i}.\mathbf{S}_j$ for a spin pair is a function of bond distance and bond angle. Atomic displacements result in changes in lattice energy $\delta E_L \sim ½\ k\ (\delta x^2 + \delta y^2 + \delta z^2)^{1/2}$ for a bond pair while the spin contribution for a given pair changes as $dE_J \sim (J(x+\delta x, y+\delta y, z+\delta z, \theta+ \delta\theta) - J(x, y, z, \theta))\ \mathbf{S_i}.\mathbf{S}_j$. The distorted structure at a given temperature will minimize the sum of spin ($E_J$ ) and lattice ($E_L$) contribution to the total energy. Strongly fluctuating local distortions will produce a random spin state while periodic local order will lead to ordered spin states. In this way, local coherent distortions which support a finite polarization are intimately coupled with the magnetic ordering. In the other direction, the magnetic order will be sensitive to the details of the local structures. External perturbations such as pressure, strain and electrical fields, for example, will affect the magnetic ground state. Indeed a rather complex magnetic ground state is found [3,5] and both pressure and substrate strain modify the magnetic properties [8,11,12].

Analysis of the x, y and z components of the force constants (self-force constants [24]) on all atoms reveals additional anomalies which can add to our qualitative description of the structural changes. The self force constants, which indicate the force on the isolated atom with respect unit displacements, are all positive indicating that the optimized structure is stable with respect to the displacement of individual atoms. The constants (Table I) can be used to estimate changes in energy for atomic displacements. a-b plane displacements by 0.005 Å for Mn, O and Ho ions yield energy changes of approximately, 0.14, 0.14 and 0.23 meV (1,1 and 2 $cm^{-1}$). These numbers are consistent with shifts observed in Raman, infrared and optical spectra [9]. (Note the large spatial asymmetry in the force constants for Mn and O1 showing the sample is stiffer along the c-axis. This is supported by recent high pressure infrared and high pressure x-ray diffraction measurements conducted by our group [10]).

To understand the bonding and spin coupling and origin of the electric polarization, we plot the total charge density for constant-y planes at y=0.33 and y=0.0. In Fig. 8(a) we can see the Mn-O1-Ho2 (short Mn-Ho) and Mn-O2-Ho2 (long Mn-Ho) distances and similar interactions in Fig. 8(b). The interesting observation to be made about the total charge density of the room temperature structure is the significant z-axis asymmetry of the charge density of the O3 and O4 ions as well as the Ho ions. This shows the origin of the electric polarization. (The same effect has been found in the charge density plots of the high temperature paraelectric phase but with symmetry about the z-axis yielding no net polarization.) Moreover, the asymmetry in the charge density enhances the overlap between Ho ions in neighboring a-b planes (~6 Å, J5) and explains the antiferromagnetic order of Ho ions observed along the c-axis [21]. Thus the electric polarization is expected to couple with the Ho ion magnetization at low temperatures.

## IV. Summary

Local structural measurements have been performed on hexagonal $HoMnO_3$ in order to ascertain the changes in bond distances which accompany magnetic ordering transitions. The observed changes in bond distances reveal strong spin-lattice coupling. The transition from paramagnetic to the antiferromagetic (noncollinear) phase near ~70 K is dominated by changes in the a-b plane Mn-Mn bond distances. The spin rotation transition near ~40 K involves both Mn-Mn and nearest neighbor Ho-Mn interactions while the low temperature transition below 10 K involves all interactions, Mn-Mn, Ho-Mn (nearest and next nearest) and Ho-Ho correlations. The similarity in magnitude of the change in J(Mn-Mn) and J(Ho-Mn) enhances the system frustration. The structural changes are interpreted in terms of a model of competing spin order and local structural distortions. Density functional calculations are used to estimate the energies associated with ionic displacements. They also reveal asymmetric polarization of the charge density of Ho, O3 and O4 sites along the z-axis in the ferroelectric phase. This polarization facilitates coupling between Ho atoms on neighboring planes parallel to the z-axis.

## V. Acknowledgments

This work is supported by DOE Grants DE-FG02-07ER46402 (NJIT) and DE-FG02-07ER46382 (Rutgers University). Data acquisition was performed at Brookhaven National Laboratory's National Synchrotron Light Source (NSLS) which is funded by the U. S. Department of Energy.

**Table I. Self-Force Constants of Ions (eV/Å$^2$)**

| Ion Site | $k_x$ | $k_y$ | $k_z$ |
|---|---|---|---|
| Mn | 8 | 8 | 27 |
| O1 | 8 | 8 | 16 |
| O2,O3 | 7 | 7 | 9 |
| O4 | 7 | 7 | 7 |
| Ho1 | 13 | 13 | 13 |
| Ho2 | 13 | 13 | 8 |

# Figure Captions

**Fig. 1.** The crystal structure of hexagonal $HoMnO_3$ showing the $MnO_5$ bipyramids in (a) and the triangular lattice of Mn ions linked by oxygen atoms in (b)

**Fig. 2**. XAFS data at 170 K for the Mn K-edge (thin line) and Ho L3-Edge (thick line).

**Fig. 3**. Fourier transform of XAFS data taken at 170 K (radial structure functions) for structure about the Mn (a) and Ho (b) sites. The thin dashed thin lines are fits to the data. Peaks are shifted to lower R-values from real distances by atomic phase shifts.

**Fig. 4.** Mn-O Bond distances as a function of temperature. The predominantly out of plane average of the Mn-O1, Mn-O2 and Mn-O3 bonds is shown as the upper curve and the in-plane Mn-O4 bond in shown as the lower curve (both in panel (a)). For comparison, the heat capacity (from Ref. [5]) and the dielectric constant (from Ref. [8]) as a function of temperature are shown in panel (b).

**Fig. 5.** The average Mn-Mn bond distance and average Ho-Ho bond distance are shown as the upper and lower curves in panel (a), respectively. In panel (b) the average Ho-Mn bond distances (short and long) are shown. The long bond exhibits a drop at the Ho ordering temperature while a dip and rise are observed in the short Ho-Mn distance at the Mn re-ordering temperature and the Ho ordering temperatures.

**Fig. 6.** Average Ho-O bond distance as a function of temperature. Note the suggestion of a small "dip" in the region of $T_{SR}$.

**Fig. 7.** Exchange paths, Ji, for HoMnO3 can be grouped into in-plane Mn-Mn (J1) and Ho-Ho (J2), Mn-Ho interactions (J3S and J3L) and the more distant in-plane Mn-Mn (J4) and between plane Mn-Mn (J5) and Ho-Ho (J6). The close proximity of $T_N$ and $T_{SR}$ indicate that J1 and J3S/J3L are similar in magnitude. O, A, B and C indicate the origin and a, b, c directions, respectively. Oxygen atoms were excluded for clarity.

**Fig. 8.** Charge density plots in atomic units obtained from a DFT calculation for the a-c plane at y=0.33 (in panel (a)) and y=0.0 (in panel (b)). The origin is in the upper right hand corner and the z-direction is down. Note the asymmetry in the charge density along the z-axis for the O3, O4 and Ho atoms which produces the low temperature polarized state. The Mn-O2-Ho2 in (a) and Mn-O1-Ho1 in (b) are the shortest distance paths which enable the Mn-Ho coupling.

**Fig. 1 Tyson *et al.***

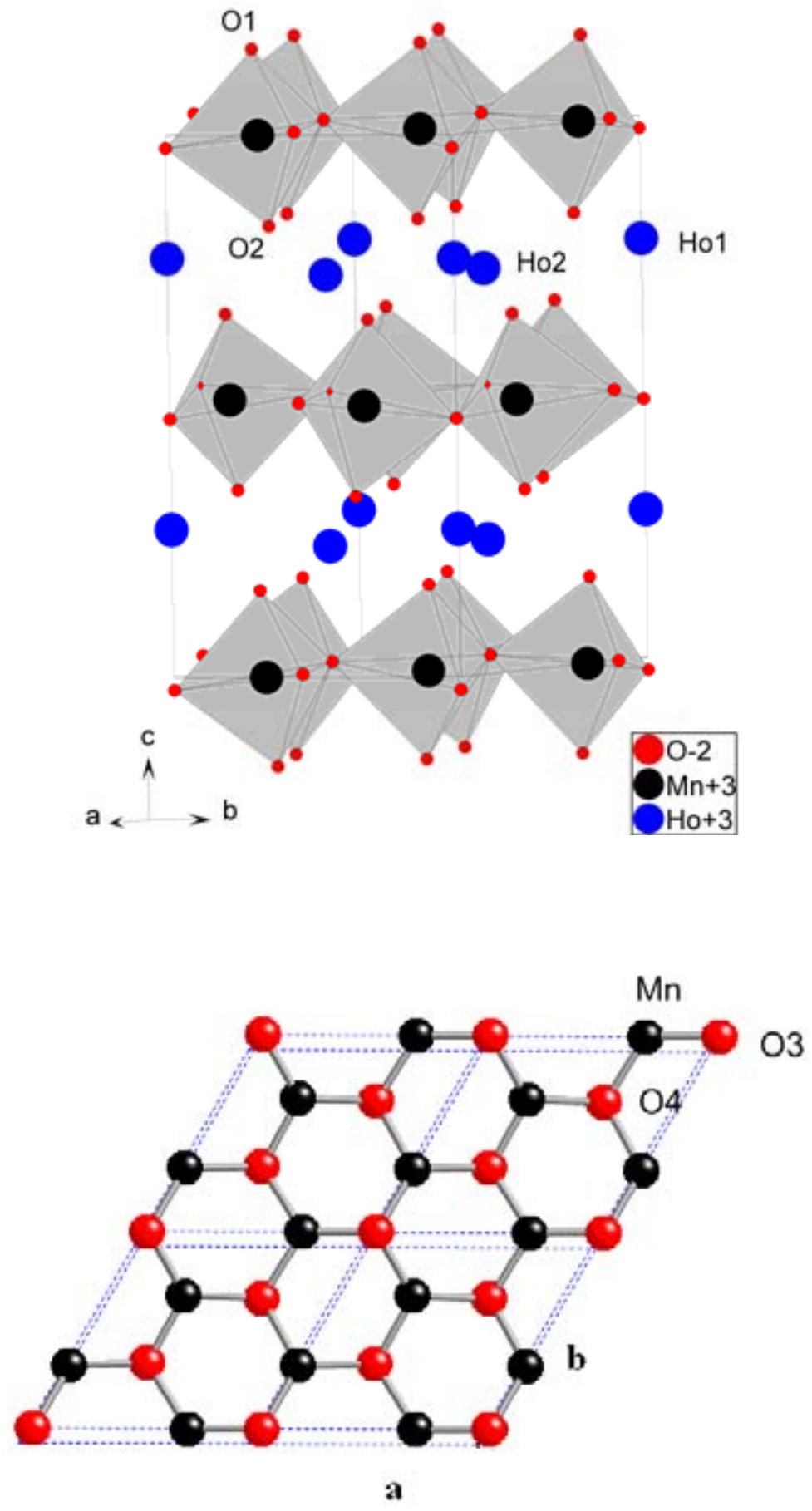

(a)

(b)

**Fig. 2 Tyson *et al.***

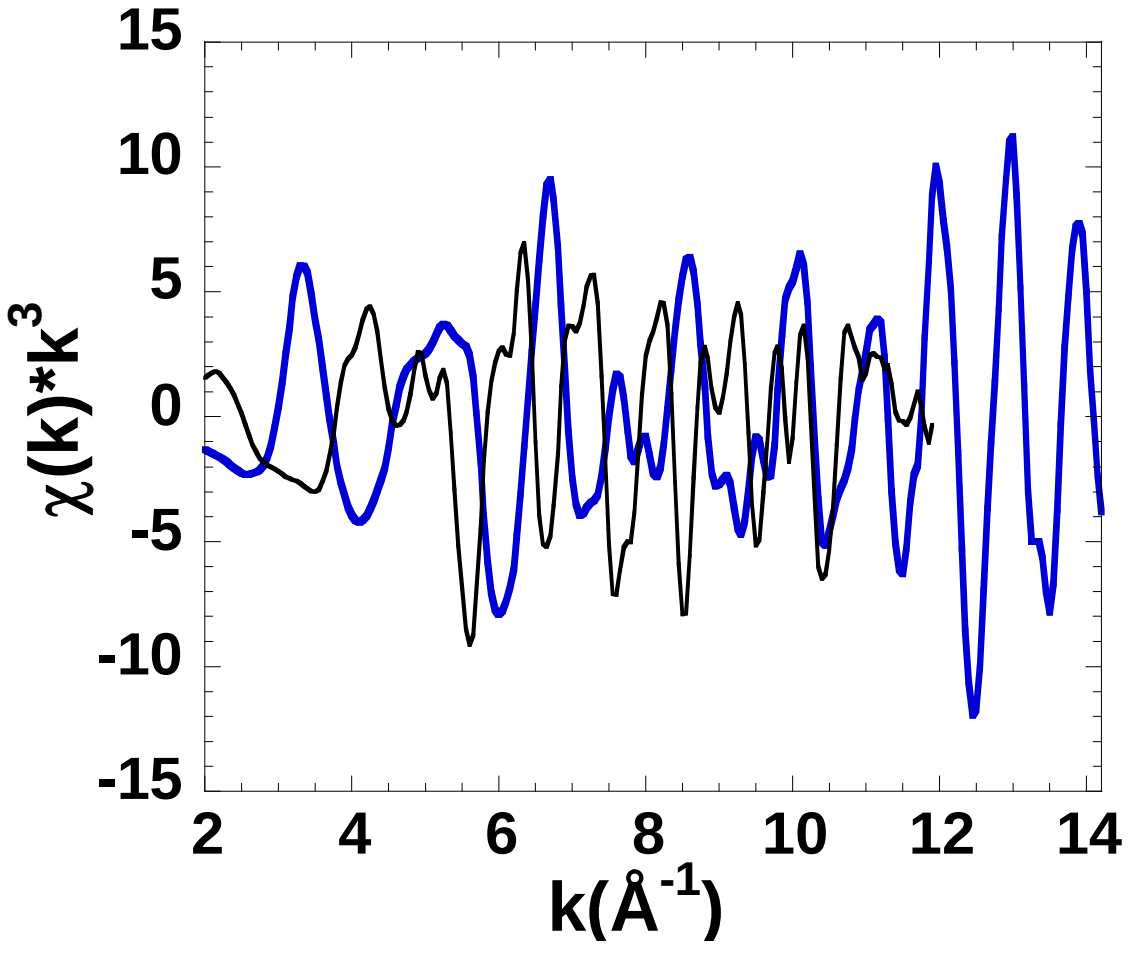

**Fig. 3   Tyson *et al.***

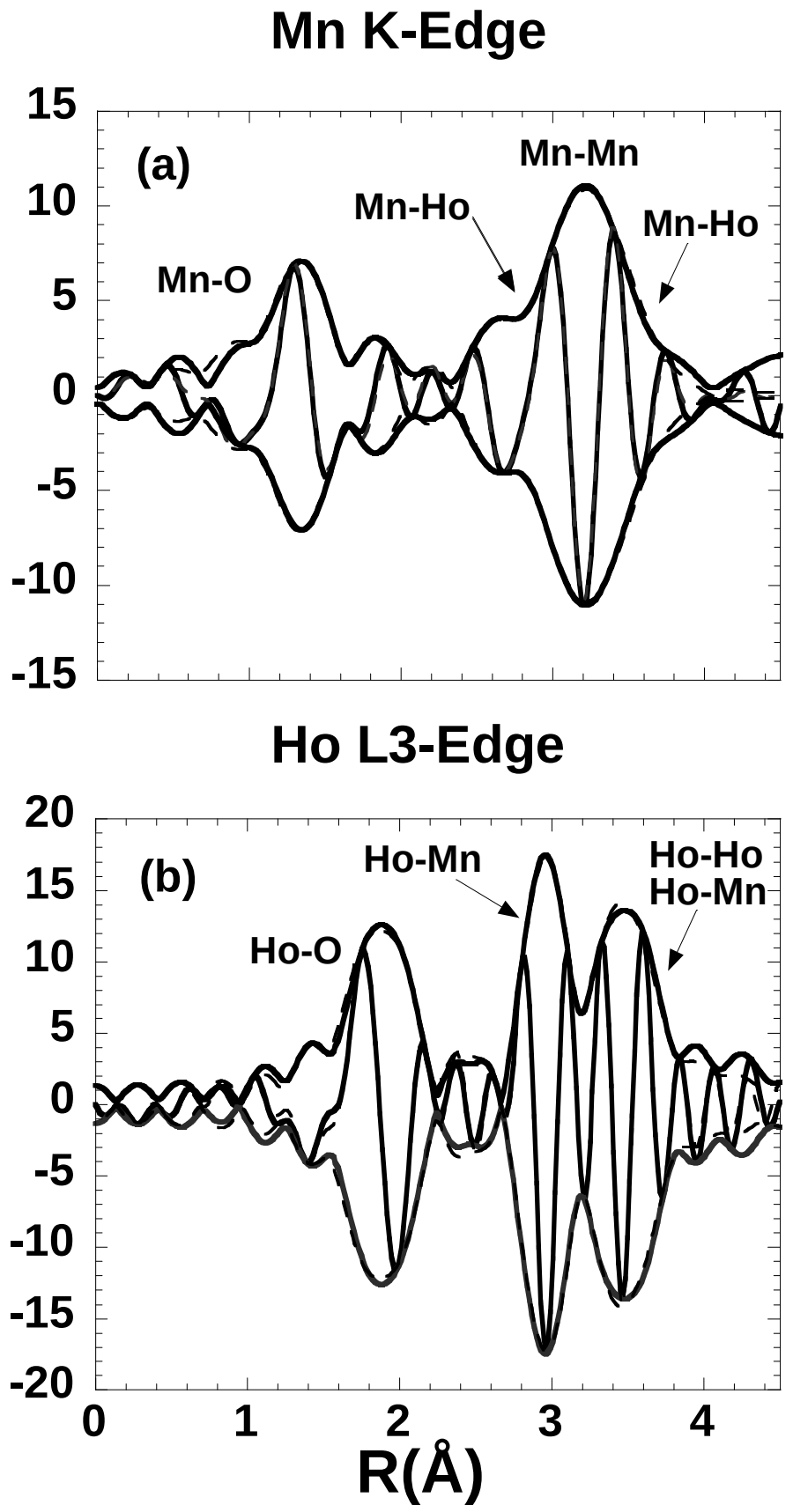

**Fig. 4  Tyson *et al.***

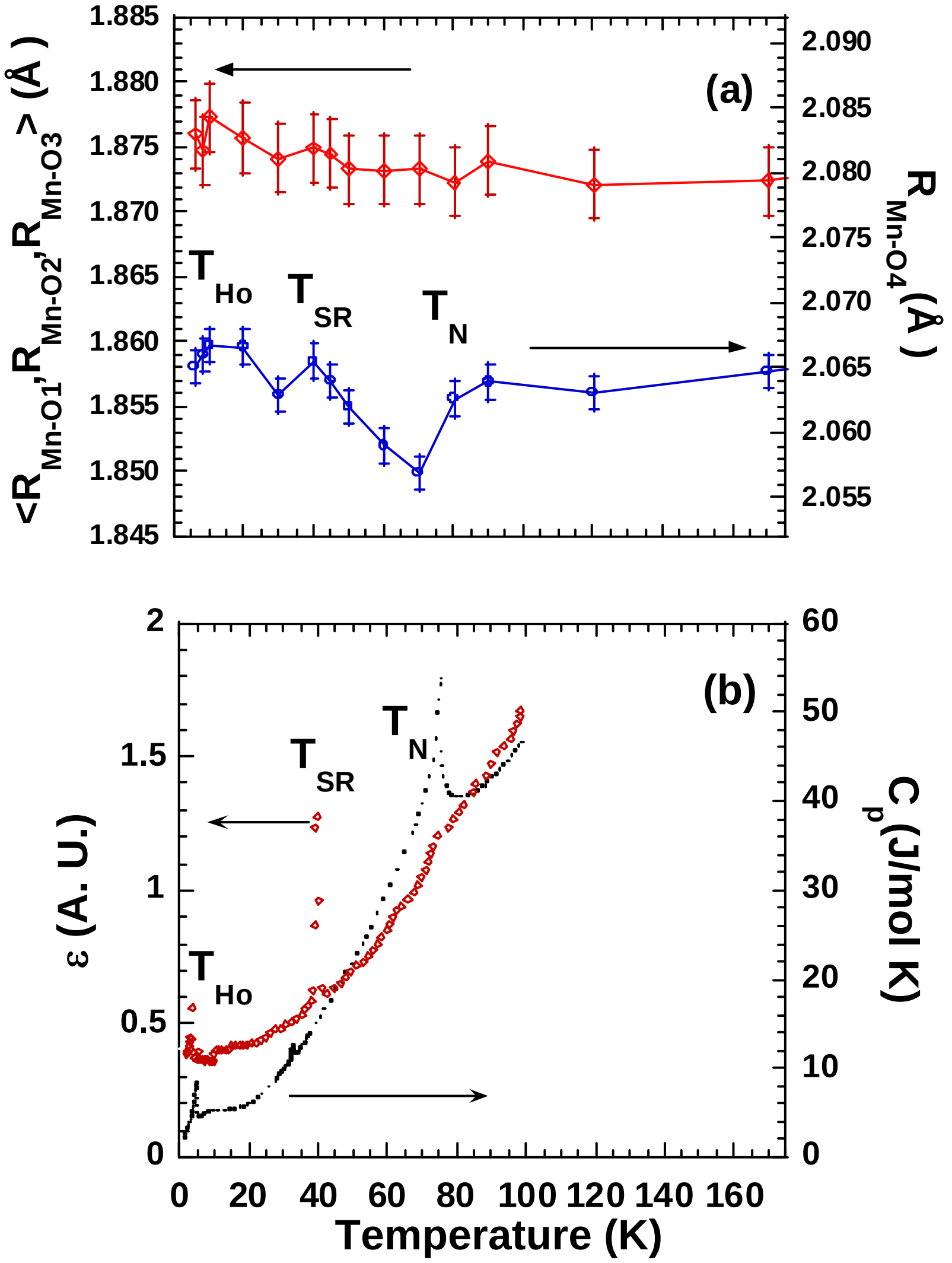

Fig. 5. Tyson *et al.*

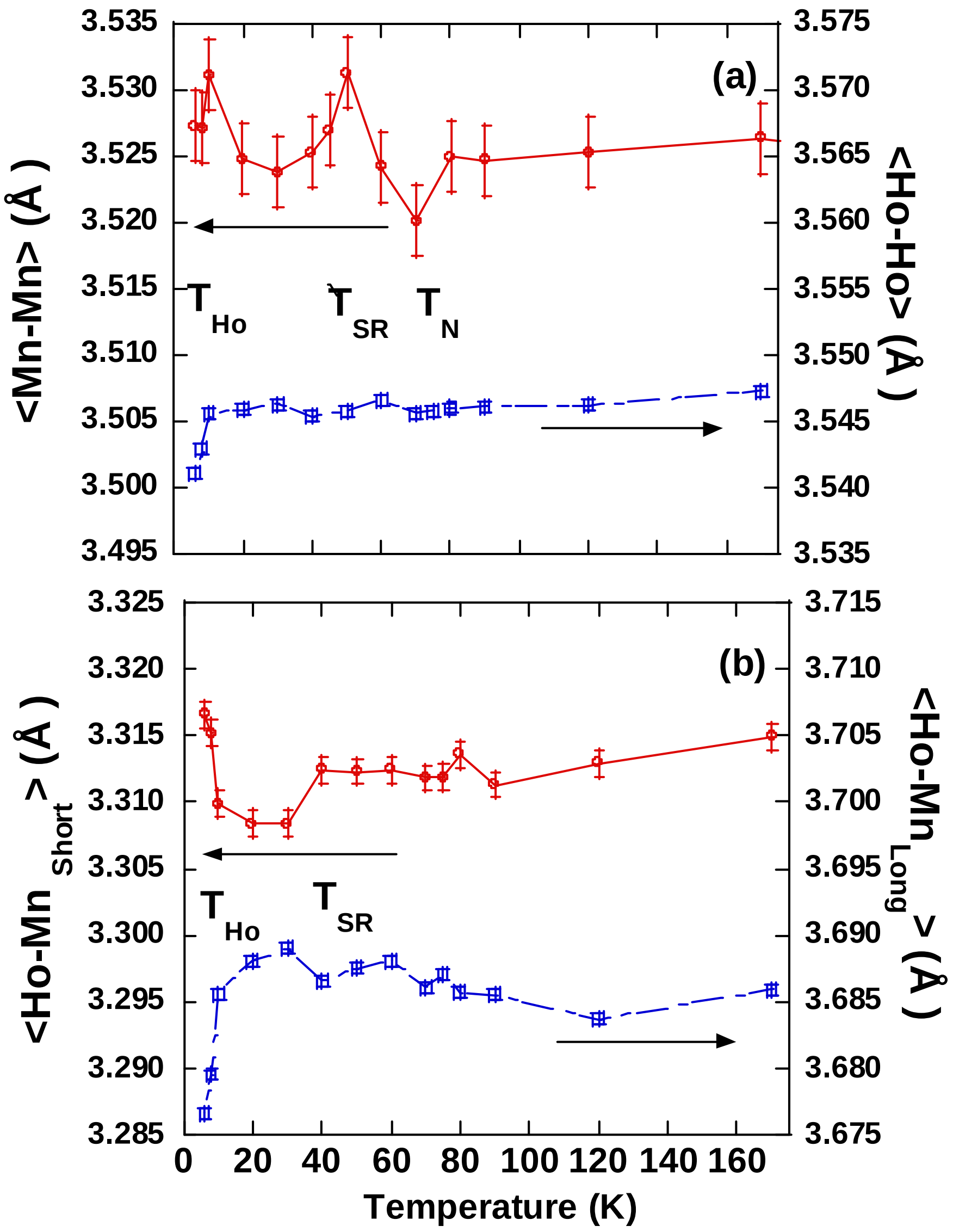

**Fig. 6. Tyson *et al.***

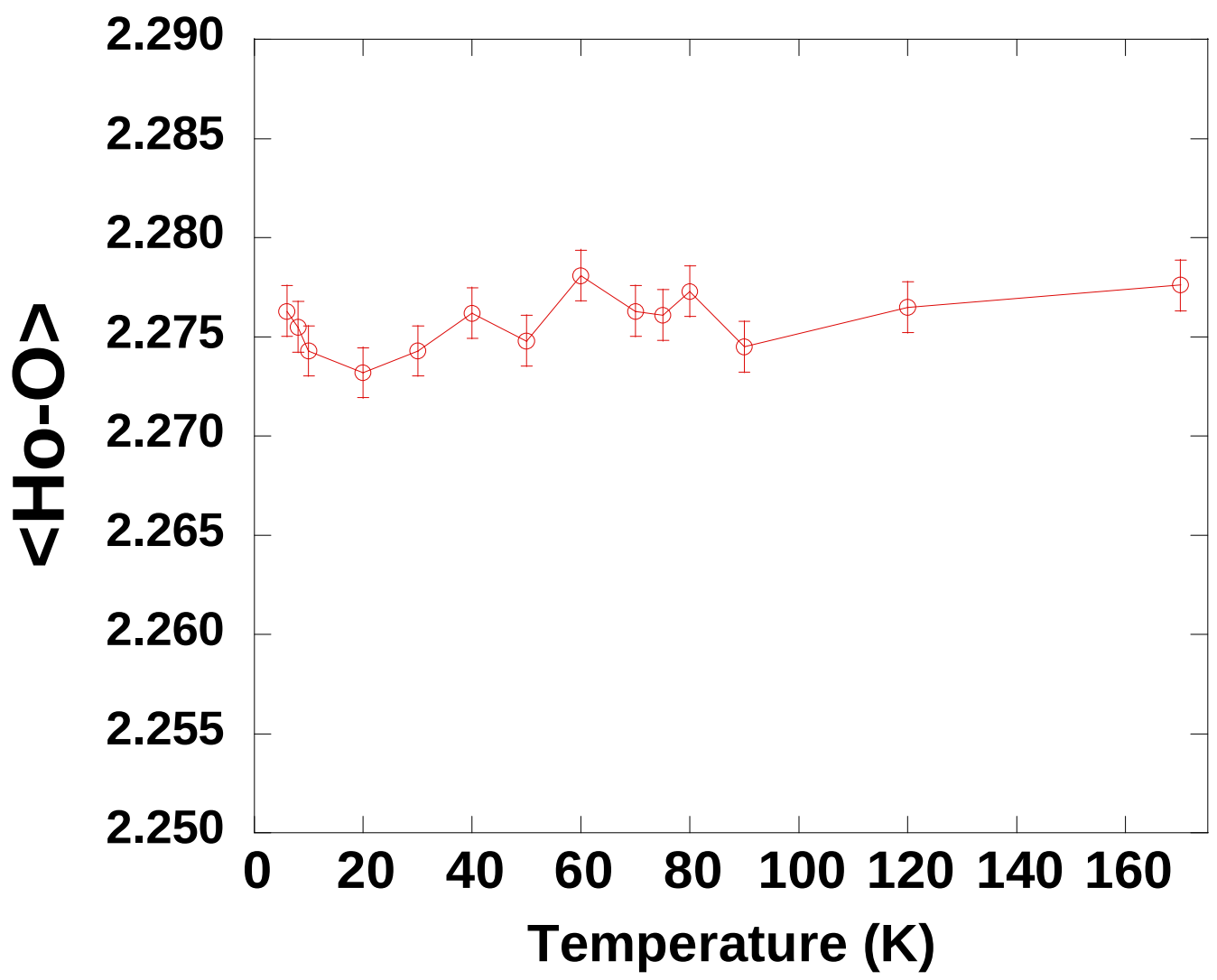

**Fig. 7. Tyson *et al.***

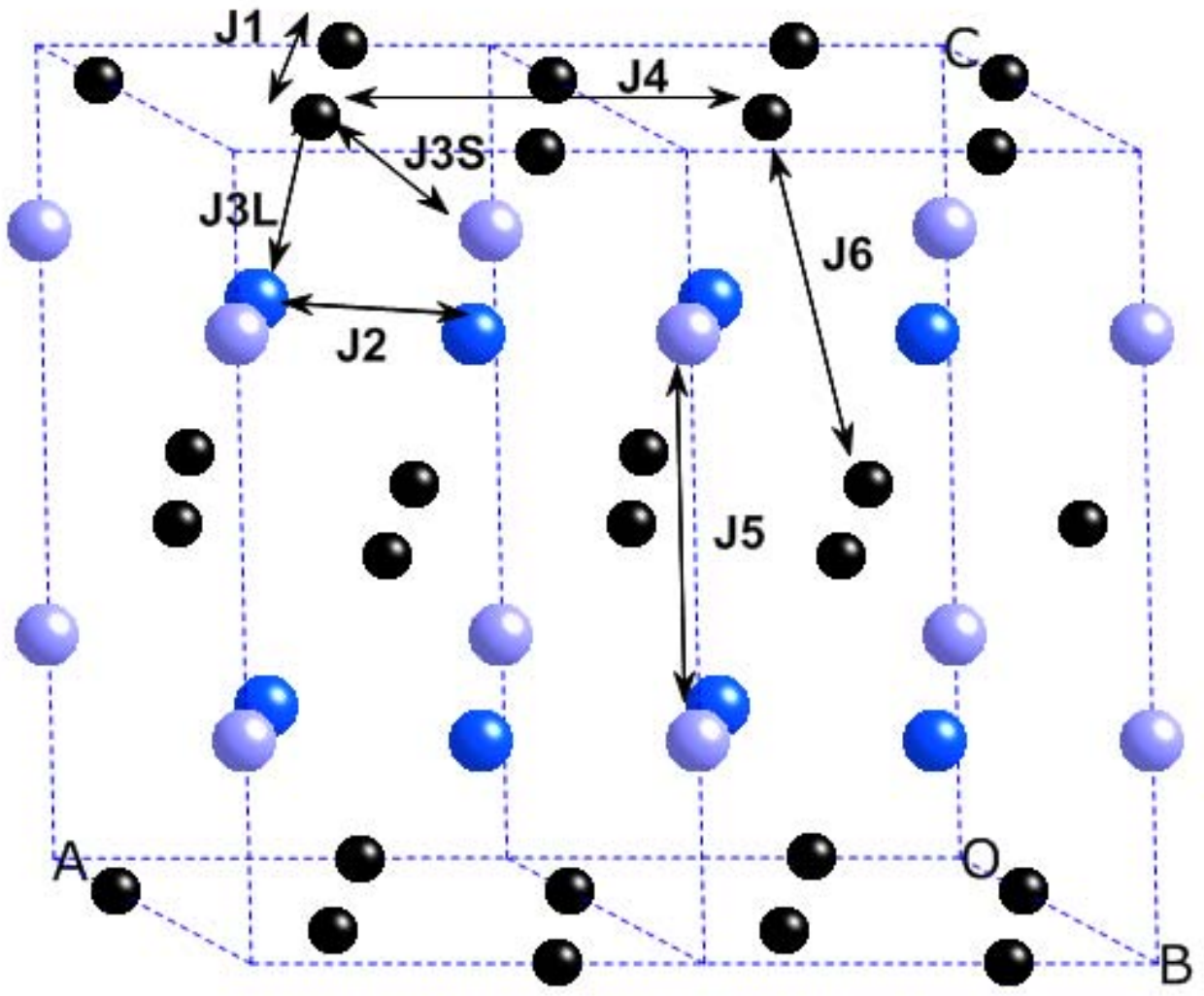

**Fig. 8. Tyson *et al.***

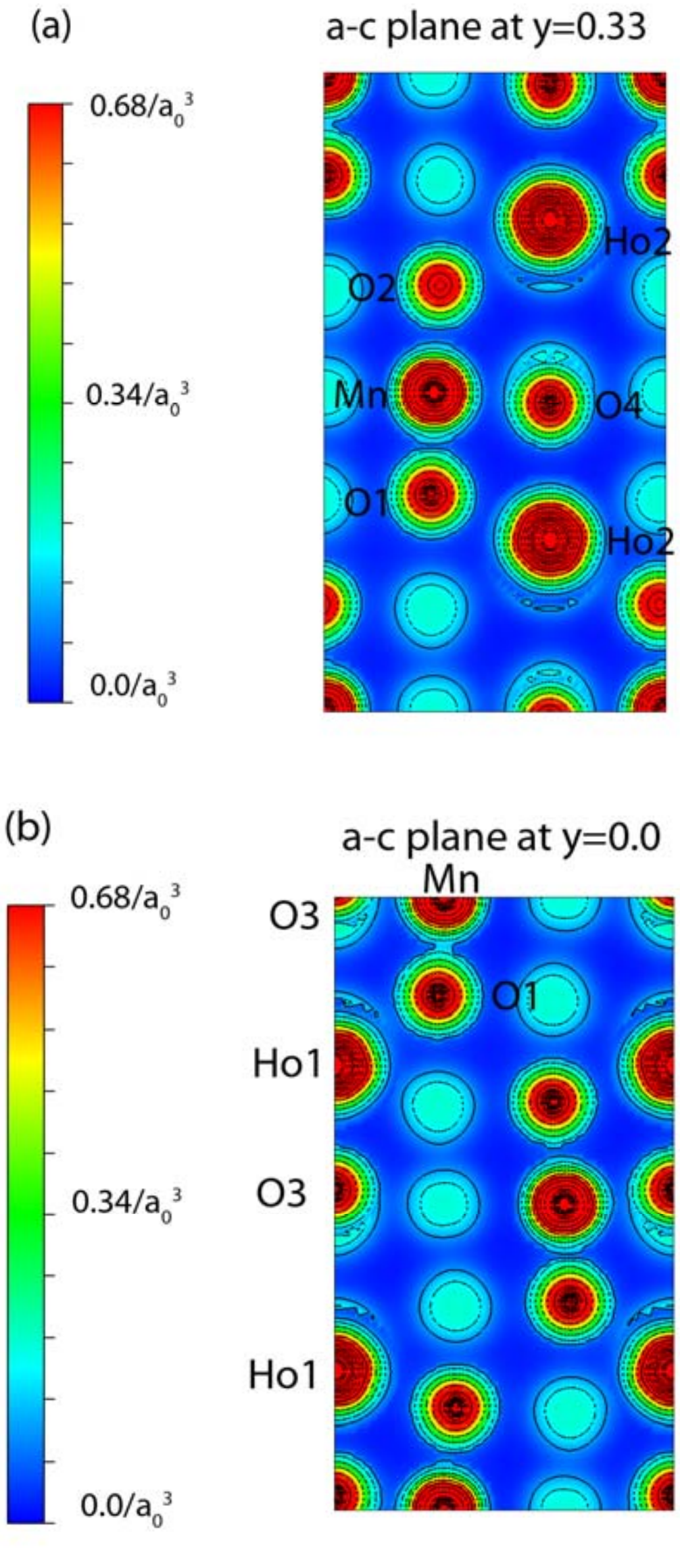